\documentclass[,final
]{aipproc}
\layoutstyle{6x9}
\usepackage{slashed}
\usepackage{ulem}
\usepackage{mathrsfs}
\usepackage{slashed}
\usepackage{amsmath}

\newcommand\gr[1]{\mathrm{#1}}
\newcommand{\bea}{\begin{eqnarray}}
\newcommand{\eea}{\end{eqnarray}}

\begin{document}
\title{Chiral multicritical points driven by isospin density in the
Ginzburg--Landau approach}
\classification{12.38.Mh, 21.65.Qr}
\keywords      {QCD phase diagram, quark matter, critical point,
inhomogeneous pion condensate}
\author{Yuhei Iwata}{
  address={Department of Physics, Tokyo University of Science,
  Tokyo 162-8601, Japan
  },
}
\author{Hiroaki Abuki}{
  address={Department of Physics, Tokyo University of Science,
  Tokyo 162-8601, Japan
  },
}
\author{Katsuhiko Suzuki}{
  address={Department of Physics, Tokyo University of Science,
  Tokyo 162-8601, Japan%
  }
}

\begin{abstract}
We study how a chiral tricritical point (TCP) on QCD phase diagram 
is affected by the imbalance of up and down quark densities (isospin
 density), using the generalized Ginzburg--Landau (GL) approach. 
The resulting phase diagram near TCP shows a rich fine structure which 
includes inhomogeneities of both the chiral and the charged pion
 condensations.  
It turns out that the TCP splits into multicritical points.
\end{abstract}

\maketitle

\section{Introduction}
Study of the phase structures of QCD at finite temperature and quark
density is intriguing subject.
In particular, it is suggested there exists a critical end point (CEP)
in the 2-flavor QCD phase diagram \cite{Asakawa:1989bq}, which leads to
a tricritical point (TCP) in the chiral limit, although the location of
the CEP is not well determined.  
According to recent work based on the Ginzburg--Landau (GL) approach
\cite{Nickel2009a,Abuki:2011pf} and effective models \cite{Nickel2009b}
for the isospin symmetric 2-flavor quark matter, the TCP is shown to be
replaced with the {\it Lifshitz} point, at which three different phases
meet; the Wigner phase, the chiral symmetry broken phase with the
spatially homogeneous chiral condensate (HCC), and the broken phase with
the inhomogeneous chiral condensate (ICC).  

How the phase structures near the TCP will change under realistic
situations for the hadronic  matter? We focus on the effect of the
isospin asymmetry, which may be realized in the quark matter under the
charge neutrality condition or  in the realistic heavy-ion collisions.
There are only a few model-based works on the TCP with the isospin
asymmetry \cite{Klein:2003fy}.  
On the other hand, the charged pion condensed phase is known to show up
as the true ground state of QCD at large isospin density \cite{Son:2000xc},
We therefore try to present a systematic model-independent approach
based on the generalized GL method to study effects of the finite
isospin density on the TCP and its neighborhood.  

In this report we focus on the  
phase structure in the vicinity of TCP for the massless $u$ and $d$ 
quark matter based on our recent work \cite{Iwata:2012bs}.  
Taking into account isospin dependent contributions to the GL
functional, we obtain new fine structures near TCP.  
As a result, the charged pion condensates extend over sizable domains in
the GL parameter space. The finite isospin density also brings about a
splitting of TCP into four independent critical points, which will be
shown below.  


\section{symmetry breaking due to the isospin density}

In the vicinity of the chiral TCP, the local free energy for the chiral
phase transition can be expressed in terms of the chiral 4-vector
$\phi=(\sigma,\pi^i)$ as order parameters, where $\sigma$ is
proportional to the chiral condensate $\langle \bar q q\rangle$ and
$\pi^i$ the pion condensates $\langle \bar qi\gamma^5\tau^iq\rangle$
$\,(i=1, 2, 3)$ with doublet quark fields $q=(u, d)$.   
In order to provide a minimal description of TCP,  
we expand the GL potential up to the sixth order of the 
order parameters, and their spatial derivatives as follows,
\begin{equation}
\begin{array}{rcl}
\Omega_{\rm GL}[\sigma({\bf x}), \pi^i({\bf x})]\!&=&\!%
\dfrac{\alpha_2}{2}\phi^2 + \dfrac{\alpha_4}{4}(\phi^2)^2 
+\dfrac{\alpha_{4b}}{4} (\nabla\phi)^2 
+\dfrac{\alpha_6}{6}(\phi^2)^3
+\dfrac{\alpha_{6b}}{6}(\phi, \nabla\phi)^2\\[2ex]
\!&&\!+\dfrac{\alpha_{6c}}{6}[\phi^2(\nabla\phi)^2-(\phi,
\nabla\phi)^2] +
\dfrac{\alpha_{6d}}{6}(\Delta\phi)^2
+ \delta\omega_c + \delta\omega_I\,,
\end{array}
\label{eq:GL}
\end{equation}
where $\alpha_i, \beta _i$ are the GL parameters, $\delta\omega_c$
stands for the contribution from current quark mass $m_c$, and
$\delta\omega_I$ expresses the contribution from the finite isospin
density. Eq.~(\ref{eq:GL}) except $\delta \omega_c$ and $\delta \omega
_I$ is invariant under the chiral $\gr{SU(2)}_L\times\gr{SU(2)}_R$
transformation. The explicit chiral symmetry breaking term, $\delta
\omega_c$, may be given by $-h\sigma$ with a small constant $h$. We set
$h = 0$ throughout this report.

For the isospin symmetry breaking term, $\delta \omega _I$, we consider the 
GL potential written with the charged pion condensates, $\pi_c^\alpha = (\pi^1,\pi^2)$, 
which  is given by 
\bea
\delta\omega_I =
\frac{\beta_2}{2}(\pi_c^\alpha)^2+\frac{\beta_4}{4}(\pi_c^\alpha)^4%
+\frac{\beta_{4b}}{4}[\phi^2-(\pi_c^\alpha)^2](\pi_c^\alpha)^2%
+\frac{\beta_{4c}}{4}(\nabla\pi_c^\alpha)^2\, .
\label{eq:GL2}
\eea
Here, we retain the $\pi_c^4$ contributions in addition to the leading 
$\pi_c^2$ terms, 
because we apply our GL analysis to the region where 
the order parameters become comparable with isospin density. 
Without loss of generality, we set $\pi^1=\pi$, $\pi^2=\pi^3=0$ from
symmetry.

The GL potential Eq.~(\ref{eq:GL}), (\ref{eq:GL2}) contains  
parameters $ \alpha_i, \beta_i$, which are, in principle, 
the function of temperature and quark/isospin chemical potentials.
In order to reduce the number of the independent parameters in 
Eq.~(\ref{eq:GL}), (\ref{eq:GL2}), we make use of the feedback of the
quark loops to the potential energy, assuming that they are the dominant
contributions.
We systematically relate the GL parameters with the $n$-th 
quark loop integrals, which certainly depend on the $u$ and $d$ quark 
chemical potential, $\mu_u$ and $\mu_d$.  
Expanding the resulting integrals up to the 2nd order of the perturbative 
isospin chemical potential $\mu_I \equiv (\mu_u -  \mu_d) / 2$, we can
show the GL parameters, $\alpha(\mu_I)$, $\beta(\mu_I)$ are proportional
to $\alpha(\mu_I = 0)$ or $\mu_I ^2 \, \alpha(\mu_I = 0)$   
\cite{Iwata:2012bs}.
Hence, we finally obtain the potential:
\begin{equation}
\begin{array}{l}
\Omega_{\rm GL}(\mu_I)=\alpha_2^{(0)}\sigma^2/2 +
(\alpha_2^{(0)} - \mu_I^2\alpha_4^{(0)}/2)\pi^2/2\\[1ex]
\;\;+(\alpha_4^{(0)}+\mu_I^2\alpha_6^{(0)})\left[\sigma^4
+ \sigma^2\pi^2 + (\sigma')^2\right]/4 + \alpha_4^{(0)}\left[\pi^4 + \sigma^2\pi^2 + (\pi')^2\right]/4\\[1ex]
\;\;+\alpha_6^{(0)}\bigl[(\sigma^2+\pi^2)^3+5(\sigma\sigma' +
\pi\pi')^2 + 3(\sigma\pi' - \pi\sigma')^2   +[(\sigma'')^2
+(\pi'')^2]/2\bigr]/6\,,
\end{array}
\label{eq:GLfd}
\end{equation}
where $\alpha^{(0)} \equiv \alpha(\mu_I=0)\,$. We restrict the analysis
 to one-dimensional structures and accordingly a primed quantity in the
 above formula means its derivative with respect to $z$-direction, for
 instance, $\sigma^\prime=\partial_z\sigma$.
We simply choose $\alpha_6^{(0)} = 1$ to set the energy scale.  
By virtue of the appropriate scaling of the GL parameters
 \cite{Abuki:2011pf}, the potential Eq.~(\ref{eq:GLfd}) is characterized
 by two independent parameters, $\{\alpha_2/\alpha_4^2,
 \mu_I^2/|\alpha_4| \}$.   
Hence, we shall describe the phase diagram in  the $(\alpha_2^{(0)},
 \alpha_4^{(0)})$ parameter space with the given 
$\mu_I$. Hereafter, we suppress the subscript $(0)$ for $\alpha$s to
 avoid notational confusion.

\section{possible phase structure at finite isospin density}

\begin{figure}[t]
\centerline{
\includegraphics[width=0.39\textwidth]{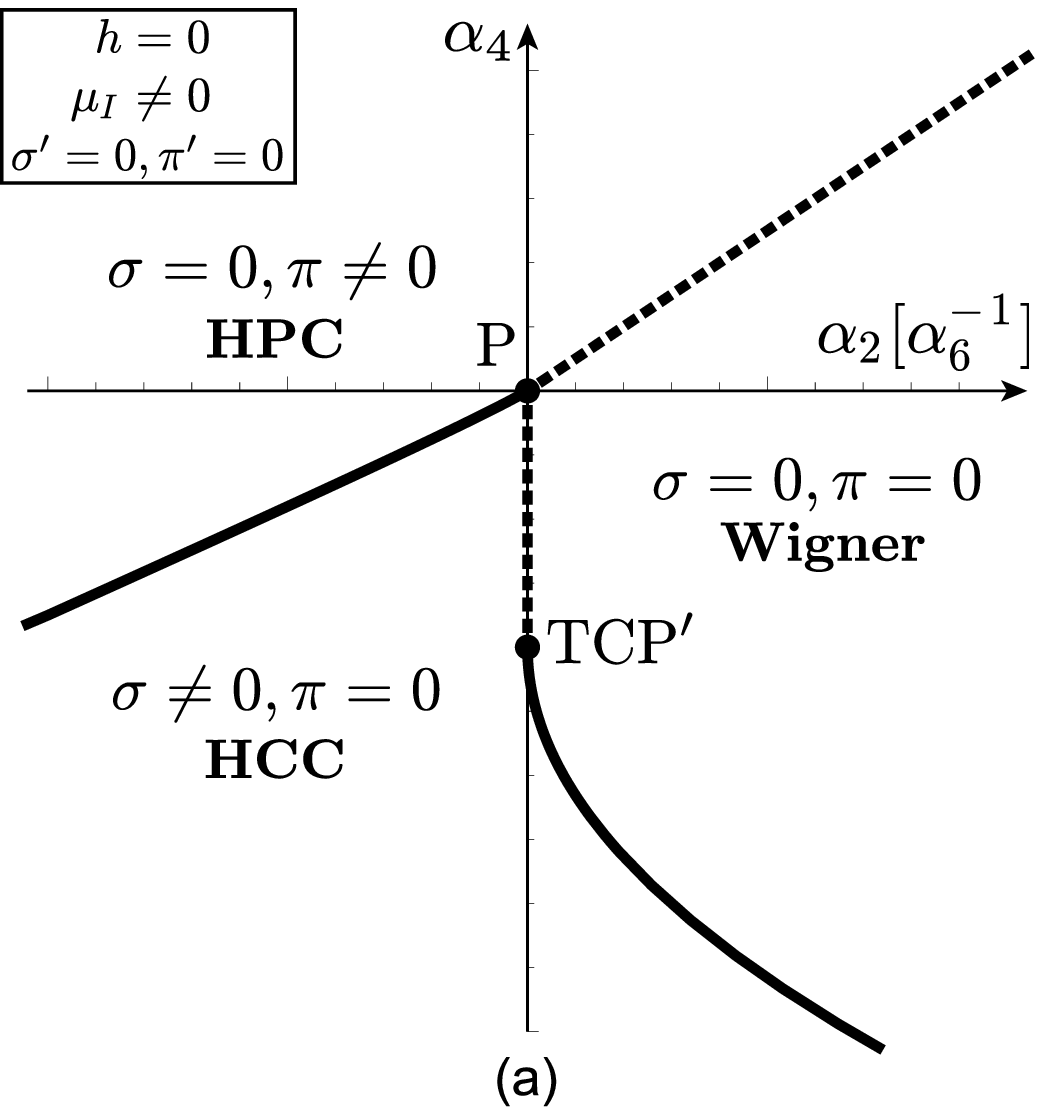}\hspace{15mm}
\includegraphics[width=0.39\textwidth]{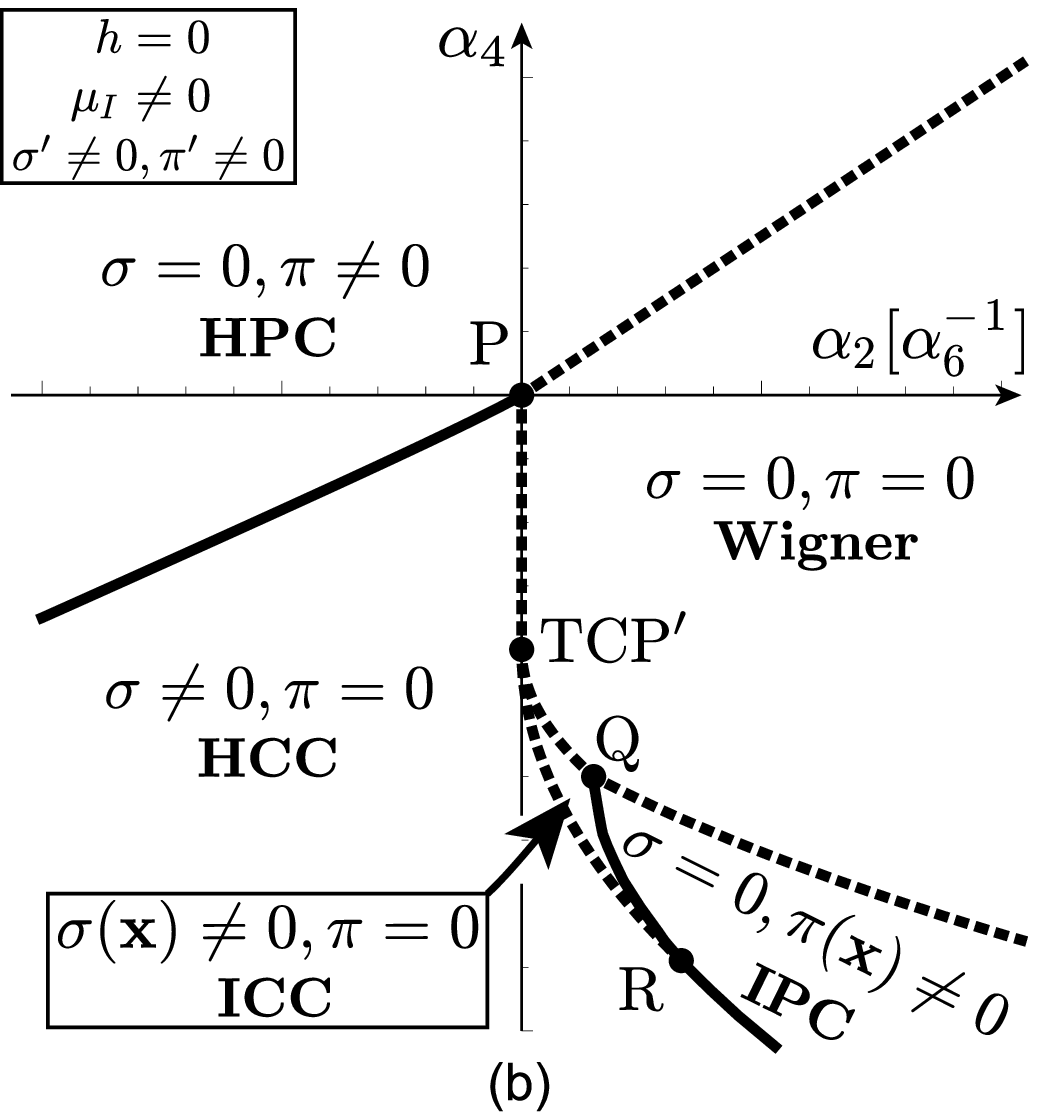}
}
\caption{%
Phase structure near the TCP for $\mu_I\ne0$, $h = 0$.  The solid and
 dashed curves indicate boundaries of the first and second order phase
 transitions, respectively.  
(a) Phase diagram with the HCC and HPC: Asymptotic behaviors of the first
 order critical line separating the HCC phase
 from the HPC phase is analytically derived:
 $\alpha_2\to 3\mu_I^2\alpha_4/4 + \mu_I^4/24 +\mathcal{O}(\mu_I^6)$ as
 $\alpha_4 \to -\infty$, while $\alpha_2 \to \mu_I^2\alpha_4/2 +
 \alpha_4^2/8$ as $\alpha_4\to 0$.
(b) Phase diagram with the ICC and IPC in addition to homogeneous condensates:
The second order critical line connecting TCP${}^\prime$ to Q is
 characterized by $\alpha_2=3(\alpha_4+\mu_I^2)^2/8$. The second order
 critical line linking TCP${}^\prime$ to R is given by $\alpha_2 = 5
 (\alpha_4 + \mu_I^2)^2/36$. The first order critical line between the
 ICC and IPC phases (a curve QR) is obtained by a numerical computation.
}
\label{fig:chirallim}
\end{figure}
 \begin{table}[t]
     \caption{The location of chiral multicritical points;
  TCP${}^\prime$ and P in Fig.~\ref{fig:chirallim}(a), Q, R in
  Fig.~\ref{fig:chirallim}(b).
 }
 \begin{tabular}{cccl}
 \hline
     & $\alpha_2$    
     & $\alpha_4$     
     & type \\ \hline
 TCP${}^\prime$ 
     & $0$
     & $-\mu_I^2$     
     & Lifshitz tricritical point \\
 P   & $0$
     & $0$            
     & bicritical point \\
 Q   & $3\mu_I^4/32$ 
     & $-3\mu_I^2/2$  
     & Lifshitz bicritical point \\
 R   & $0.21\mu_I^4$ 
     & $-2.22\mu_I^2$ 
     & critical (end) point \\
 \hline
 \end{tabular}
 \label{tab:tab1}
  \end{table}

Let us first assume  the spatially constant distribution for the chiral
and charged pion condensates.  
We show the phase diagram near the TCP in Fig.~\ref{fig:chirallim} (a),
and locations of multicritical points on the GL parameter space in TABLE
\ref{tab:tab1}. In this case, the ground state favors either HCC phase
or homogeneous pion condensed (HPC) phase. There is no room for the
coexistence phase of $\pi$ and $\sigma$. Compared with the result for
the $\mu_I = 0$  matter, large domains of the phase diagram is occupied
by the HPC phase. This drastic change is due to the presence of
$(\alpha_2 - \mu_I^2\alpha_4/2) \pi^2 / 2$ term in the GL potential
Eq.~\eqref{eq:GLfd}, which favors the HPC phase for arbitrary small
$\mu_I$ with $\alpha_ 4 > 0$. In addition, while the TCP is located at
the origin of the phase diagram for the $\mu_I=0$  matter, 
the finite  $\mu_I$ brings about a shift of the location of TCP to $(0,
-\mu^2_I)$, newly labeled by TCP$'$. Origin of this shift is easily
understood by looking at the coefficient of the $\sigma^4$ term in
Eq.\eqref{eq:GLfd}, $\alpha_4^{(0)} + \mu_I^2 \alpha_6$, $(\alpha_6 =
1)$. Our result is consistent with the model-based calculations
\cite{Klein:2003fy}. 

Next, we consider possible inhomogeneous structures for the chiral and 
charged pion condensates with the finite isospin chemical potential.   
Following  the previous work, we consider the one-dimensional solitonic
structure expressed by the Jacobi's elliptic function ``sn'', for the
inhomogeneous phases of $\sigma$ and $\pi$ condensates
\cite{Abuki:2011pf}. In fact, the Jacobi's elliptic function is shown to
be a solution of the Euler-Lagrange equation for the GL potential in the
chiral limit. We assume $\sigma({\bf x})  = \sqrt{\nu} k\,{\rm{sn}}{(kz;
\nu)}$, $\pi = 0$ for the ICC phase, and $\sigma = 0$, $\pi ({\bf x}) =
\sqrt{\nu} k\,{\rm{sn}}{(kz; \nu)}$ for inhomogeneous charged pion
condensed (IPC) phase, where the positive parameters $\nu$ and $k$ are
determined by the variational principle in each phase. The resulting
phase diagram is depicted in Fig.~\ref{fig:chirallim} (b).

The TCP is now replaced with the {\it Lifshitz} tricritical point.
Compared with isospin symmetric matter, a large part of the ICC phase is
taken over by the IPC. This is because the spatial derivative term of
$(\pi')^2$ gets lower energy compared with the  $(\sigma')^2$
contribution as clearly seen in Eq.~\eqref{eq:GLfd}. The boundary
between ICC and IPC also provides two new critical points, Q and R,
listed in Table \ref{tab:tab1}.  

For completeness, we also consider possible $\sigma$-$\pi$ coexistence
phases. We take the ansatz of {\it chiral kink spiral}: $\sigma({\bf
x})=m\,{\rm{cn}}{(kz; \nu)}, \pi({\bf x}) = m\,{\rm{sn}}{(kz;\nu)}$
with $m$ a new variational parameter. This includes the standard chiral
spiral $\sigma({\bf x})=m\cos{(kz)}, \pi({\bf x})=m\sin{(kz)}$
\cite{Nakano:2004cd} in a particular limit $\nu\to 0$. We always find
either IPC or ICC being favored, indicating that the $\sigma$-$\pi$
coexistence state may be ruled out near TCP.

To summarize, we studied the phase structures in the proximity of TCP at
finite $\mu_I$, using the model-independent GL approach. The sizable
domain of the phase diagram was found to be occupied by the charged pion
condensation. We found not only the shift of the chiral TCP, but also
the appearance of the new critical points. The new fine structure
obtained in the GL parameter space may be mapped onto the low
temperature and intermediate density of the $(T,\mu)$ QCD phase diagram,
as suggested in the previous work \cite{Nickel2009b}. The inhomogeneous
charged pion condensate of the quark matter may smoothly continue to the
classical pion condensate discussed in the nuclear matter
\cite{Migdal:1990vm}.

\bibliographystyle{aipproc}

\end{document}